\documentclass[fleqn,usenatbib,useAMS]{mnras}

\usepackage{mathptmx}

\usepackage[T1]{fontenc}
\usepackage{ae,aecompl}

\usepackage{graphicx}
\usepackage{subcaption}
\usepackage[section]{placeins}
\DeclareGraphicsExtensions{.pdf,.png,.jpg,.eps}
\usepackage{epstopdf}
\usepackage{color}
\usepackage{amsmath}
\usepackage{amssymb}
\usepackage{amsbsy}
\usepackage{comment}
\usepackage[normalem]{ulem}
\usepackage{enumitem}

\newcommand{\green}[1]{\iffalse #1 \fi}
\usepackage{pbox}

\definecolor{blue}{rgb}{0,0,1}

\usepackage{graphicx}	
\usepackage{amsmath}	
\usepackage{amssymb}	
\usepackage{bm}		
\usepackage{pdflscape}	
\usepackage{xcolor}
\usepackage{exscale}
\usepackage{natbib}
\usepackage{rotating}
\usepackage{rotate}
\usepackage{afterpage}
\usepackage{booktabs,threeparttable}
\usepackage{relsize}
\usepackage{lscape}
\usepackage{tabularx}
\usepackage{longtable}
\usepackage{ltxtable}
\usepackage{txfonts}
\usepackage{bbm}
\usepackage{afterpage}
\usepackage{bbold}
\usepackage{footnote}
\usepackage{widetext}

\title[Galaxy Clustering in Harmonic Space from DES-Y1]{Galaxy Clustering in Harmonic Space from the Dark Energy Survey Year 1 Data: Compatibility with Real-Space Results}

\author[DES Collaboration]{
\parbox{\textwidth}{
  \Large
F.~Andrade-Oliveira,$^{1,2 \dagger}$
H.~Camacho,$^{1,2}$
L.~Faga,$^{3,2}$
R.~Gomes,$^{3,2}$
R.~Rosenfeld,$^{4,2}$
A.~Troja,$^{4,2}$
O.~Alves,$^{5,1,2}$
C.~Doux,$^{6}$
J.~Elvin-Poole,$^{7,8}$
X.~Fang,$^{9}$
N.~Kokron,$^{10,11}$
M.~Lima,$^{3,2}$
V.~Miranda,$^{9}$
S.~Pandey,$^{6}$
A.~Porredon,$^{7,12,13}$
J.~Sanchez,$^{14}$
M.~Aguena,$^{3,2}$
S.~Allam,$^{14}$
J.~Annis,$^{14}$
S.~Avila,$^{15}$
E.~Bertin,$^{16,17}$
D.~Brooks,$^{18}$
D.~L.~Burke,$^{11,19}$
M.~Carrasco~Kind,$^{20,21}$
J.~Carretero,$^{22}$
R.~Cawthon,$^{23}$
C.~Chang,$^{24,25}$
A.~Choi,$^{7}$
M.~Costanzi,$^{26,27,28}$
M.~Crocce,$^{12,13}$
L.~N.~da Costa,$^{2,29}$
M.~E.~S.~Pereira,$^{5}$
S.~Desai,$^{30}$
H.~T.~Diehl,$^{14}$
P.~Doel,$^{18}$
A.~Drlica-Wagner,$^{24,14,25}$
S.~Everett,$^{31}$
A.~E.~Evrard,$^{32,5}$
I.~Ferrero,$^{33}$
J.~Frieman,$^{14,25}$
J.~Garc\'ia-Bellido,$^{15}$
E.~Gaztanaga,$^{12,13}$
D.~W.~Gerdes,$^{32,5}$
D.~Gruen,$^{10,11,19}$
R.~A.~Gruendl,$^{20,21}$
S.~R.~Hinton,$^{34}$
D.~L.~Hollowood,$^{31}$
B.~Jain,$^{6}$
D.~J.~James,$^{35}$
N.~Kuropatkin,$^{14}$
O.~Lahav,$^{18}$
N.~MacCrann,$^{36}$
M.~A.~G.~Maia,$^{2,29}$
P.~Melchior,$^{37}$
F.~Menanteau,$^{20,21}$
R.~Miquel,$^{38,22}$
R.~Morgan,$^{23}$
J.~Myles,$^{10,11,19}$
R.~L.~C.~Ogando,$^{2,29}$
A.~Palmese,$^{14,25}$
F.~Paz-Chinch\'{o}n,$^{20,39}$
A.~A.~Plazas~Malag\'on,$^{37}$
M.~Rodriguez-Monroy,$^{40}$
E.~Sanchez,$^{40}$
V.~Scarpine,$^{14}$
S.~Serrano,$^{12,13}$
I.~Sevilla-Noarbe,$^{40}$
M.~Smith,$^{41}$
M.~Soares-Santos,$^{5}$
E.~Suchyta,$^{42}$
G.~Tarle,$^{5}$
and C.~To$^{10,11,19}$
\begin{center} (DES Collaboration) \end{center}
}
\vspace{-1cm}
\\
}

\newcommand{\eg}{e.g$.$ }

\newcommand{\ie}{i.e$.$ }

\newcommand{\nv}{\hat{\bf n}}

\definecolor{ForestGreen}{rgb}{0.133, 0.545, 0.133}

\usepackage[mathlines]{lineno}

\usepackage{etoolbox} 

\newcommand*\linenomathpatch[1]{%
  \expandafter\pretocmd\csname #1\endcsname {\linenomath}{}{}%
  \expandafter\pretocmd\csname #1*\endcsname{\linenomath}{}{}%
  \expandafter\apptocmd\csname end#1\endcsname {\endlinenomath}{}{}%
  \expandafter\apptocmd\csname end#1*\endcsname{\endlinenomath}{}{}%
}
\newcommand*\linenomathpatchAMS[1]{%
  \expandafter\pretocmd\csname #1\endcsname {\linenomathAMS}{}{}%
  \expandafter\pretocmd\csname #1*\endcsname{\linenomathAMS}{}{}%
  \expandafter\apptocmd\csname end#1\endcsname {\endlinenomath}{}{}%
  \expandafter\apptocmd\csname end#1*\endcsname{\endlinenomath}{}{}%
}

\expandafter\ifx\linenomath\linenomathWithnumbers
  \let\linenomathAMS\linenomathWithnumbers
  \patchcmd\linenomathAMS{\advance\postdisplaypenalty\linenopenalty}{}{}{}
\else
  \let\linenomathAMS\linenomathNonumbers
\fi

\linenomathpatch{equation} 
\linenomathpatchAMS{gather}
\linenomathpatchAMS{multline}
\linenomathpatchAMS{align}
\linenomathpatchAMS{alignat}
\linenomathpatchAMS{flalign}

\begin{document}

\maketitle
\date{\today}

\begin{abstract}
We perform an analysis in harmonic space of the  Dark Energy Survey Year 1 Data (DES-Y1) galaxy clustering data using
products obtained for the real-space analysis. 
We test our pipeline with a suite of lognormal simulations, which are used to validate scale cuts in harmonic space as well as to provide a covariance matrix that takes into account the DES-Y1 mask.
We then apply this pipeline to DES-Y1 data taking into account survey property maps derived for the real-space analysis. We compare with real-space DES-Y1 results obtained from a similar pipeline.
We show that the harmonic space analysis we develop yields results that are compatible with the real-space analysis for the bias parameters.  This verification paves the way to performing a harmonic space analysis for the upcoming DES-Y3 data. 
\end{abstract}


\makeatletter
\def \blfootnote{\xdef\@thefnmark{}\@footnotetext}
\makeatother

\blfootnote{$^{\dagger}$ E-mail: felipe.andrade-oliveira@unesp.br}

${}$

\section{Introduction}

Cosmology has matured into a precision, data-driven science in the last couple of decades. An immense amount of data from different observables, including the detailed study of the cosmic microwave background, the abundance of light elements, the detection of thousands of type Ia supernovae and the distribution of galaxies and their shapes as measured in large galaxy surveys has confirmed the standard spatially flat $\Lambda$CDM cosmological model \citep[e.g][]{Frieman:2008sn}.
There are, however, some tensions among some of these observables that, if they stand, can point to important modifications in our understanding of the universe \citep{Verde:2019ivm}. Therefore, the testing of the standard cosmological model and the search for new phenomena continue with new data and improved analysis methods. 

The main cosmological analysis of several recent galaxy surveys
uses the measurement of 2-point correlation functions (or power spectra) of observables, such as galaxy clustering and cosmic shear, as inputs to the estimation of cosmological parameters of a given model in a likelihood framework. The use of a joint combination of these 2-point correlations is conventionally called the ``3x2-point"  analysis including galaxy-galaxy, galaxy-galaxy lensing and galaxy lensing-galaxy lensing (shear) 2-point correlations.
These analyses can be performed in real-space with the angular correlation functions or in harmonic space with the angular power spectra. 
There are advantages and disadvantages in both approaches. For an idealized full sky survey, the harmonic modes are independent on linear scales and the covariance matrices are diagonal. This is not the case in real space, where the angular correlation function presents large correlations at different angular scales. For realistic surveys, however, the effect of the survey mask introduces mode coupling that makes the analyses more convoluted and the real-space measurements are in general more amenable to the presence of the mask. 
In principle the data contains the same amount of information whether it is analysed in real or harmonic space if all modes or scales are included but in reality differences may arise due to finite survey area and the different, independent methods that are used ({\it e.g.} the definition of scale cuts).

The Dark Energy Survey (DES\footnote{\tt www.darkenergysurvey.org}) used photometric redshift measurements to perform tomographic
real-space analyses of galaxy clustering \citep{Elvin-Poole:2017xsf},
cosmic shear \citep{Troxel:2017xyo} and galaxy-galaxy lensing \citep{Prat:2017goa}, culminating in a joint 
3x2-point analysis \citep{Abbott:2017wau} for its first year of data (DES-Y1). The only harmonic space analysis from DES-Y1 data so far was the study of baryon acoustic oscillations \citep{Camacho:2018mel, Abbott:2017wcz}.

Other photometric surveys have recently presented results in harmonic space. The Kilo-Degree Survey (KiDS\footnote{\tt kids.strw.leidenuniv.nl} ) has presented a harmonic- and real- space analysis of cosmic shear \citep{Kohlinger:2017sxk}
and 3x2-point analysis also in harmonic space in combination with different data sets \citep{vanUitert:2017ieu,2020arXiv200715632H}. \cite{Balaguera-Antolinez:2017dpm} investigated the clustering in harmonic space in the local Universe using the 2MASS Photometric Redshift catalogue (2MPZ) \citep{Bilicki:2013sza}.
The  Subaru Hyper Suprime-Cam (HSC\footnote{\tt hsc.mtk.nao.ac.jp/ssp}) has performed a cosmic shear analysis from its first year of data both in harmonic \citep{Hikage:2018qbn} and real-space \citep{Hamana:2019etx}. 
Recently, \citet{2020JCAP...03..044N} undertook an independent investigation of the galaxy clustering in harmonic space using HSC public data.

We should also mention that in spectroscopic surveys the clustering analyses are performed in three dimensions, since they have access to more reliable spectroscopic redshift measurements. The most recent results come from the completed Sloan Digital Sky Survey IV (SDSS-IV) extended Baryon Oscillation Spectroscopic Survey, eBOSS\footnote{\tt www.sdss.org/surveys/eboss/} \citep{2020arXiv200708991E} using both the 2-point correlation function \citep{2021MNRAS.500..736B,Tamone_2020,2021MNRAS.500.1201H} and the power spectrum \citep{2020MNRAS.498.2492G,deMattia:2020fkb,2020MNRAS.499..210N} for different tracers (Luminous Red Galaxies, Emission Line Galaxies and Quasars).
In addition, two-dimensional angular clustering analysis with SDSS data were also performed with the DR12 data, using both the angular correlation funcion~\citep{2017MNRAS.468.2938S} and the angular power spectrum~\citep{2019MNRAS.485..326L}.

We perform an analysis in harmonic space of galaxy clustering from DES-Y1 data using the galaxy sample and the survey systematic maps with the corresponding weights from the real-space analysis \citep{Elvin-Poole:2017xsf}. In order to compare the harmonic space analysis with the real space one we adopt the same fiducial cosmology used in \citet{Elvin-Poole:2017xsf}, namely a flat $\Lambda$CDM model with cosmological parameters 
$\Omega_m      =  0.276$,
$h_0           =  0.7506$,
$\Omega_b      =  0.0531$,
$n_s          =  0.9939$
$A_s          =  2.818378 \times 10^{-9}$ and
$ \Omega_\nu h^2      =  0.00553$. 
This will be referred to as DES-Y1 cosmology. In this cosmology the amplitude of perturbations is fixed at $\sigma_8 = 0.83$. Our main goal is to develop and test the tools for the harmonic space analyses of galaxy clustering, demonstrating the compatibility with the DES-Y1 results in real space.   


This paper is organized as follows. In section \ref{sec:model} we describe the theoretical modelling of the angular power spectrum, presenting in section \ref{sec:method} the pseudo-$C_\ell$ method used to measure the angular power spectrum in a masked sky. Section \ref{sec:mocks} details the generation of lognormal mocks. The results of measurements of the galaxy clustering in the mocks are shown in section \ref{sec:mockmeasurement} and different covariance matrices are compared in section \ref{sec:covariance}. The pipeline that we develop for the estimation of parameters from the angular power spectrum is presented in section \ref{sec:resultsmocks} where we discuss the adopted scale cuts and it is applied on the mocks in \ref{sec:pipelineonmocks}. Finally, section \ref{sec:resultsdata} shows our results on DES-Y1 data and we present our conclusions in section \ref{sec:conclusions}.

\section{Theoretical modelling}
\label{sec:model}
    The starting point of the modelling of the galaxy angular power spectrum is the 3D non-linear matter power spectrum $P(k,z)$ at a given wavenumber $k$ and redshift $z$. We obtain it by using either of the Boltzmann solvers CLASS \footnote{\tt www.class-code.net} or CAMB \footnote{\tt camb.info} to calculate the linear power spectrum and the HALOFIT fitting formula \citep{Smith2003} in its updated version \citep{Takahashi2012} to turn this into the late-time non-linear power spectrum.  The galaxy angular power spectrum function can then be derived from this 3D power spectrum in the Limber approximation \citep{Limber1953,LoVerde:2008re} as \citep[\eg][]{Krause2017}:
\begin{equation}
\label{eq:C_ell_delta_g}
    C^{ij}_{\delta_g \delta_g}(\ell) = \int d\chi \frac{q^i_g\left(\frac{\ell+\frac{1}{2}}{\chi},\chi \right) q^j_g\left(\frac{\ell+\frac{1}{2}}{\chi},\chi \right)}{\chi^2} P\left(\frac{\ell+\frac{1}{2}}{\chi}, z(\chi)\right)\ ,
\end{equation}
where $\chi$ is the comoving radial distance, $i$ and $j$ denote different combinations of photometric redshift bins and 
the radial weight function for clustering $q^i_g$ is given by
\begin{equation}
 q^i_g (k,\chi) = b^i(k, z(\chi))\  n^i_g(z(\chi)) \frac{d z}{d\chi}\ .
\end{equation}
Here $H_0$ is the Hubble parameter today, $\Omega_m$ the ratio of today's matter density to today's critical density of the universe, $z(\chi)$ is the redshift at comoving distance $\chi$ and $b^i(k, z)$ is a scale and redshift dependent galaxy bias. Furthermore, $n^i_{g}(z)$ denote the redshift distributions of the DES-Y1 lens galaxies, normalised such that
\begin{equation}
    \int dz \; n^i_{g}(z) = 1\ .
\end{equation}

Here we assume a simple linear bias model, constant for each redshift bin, \ie $b^i(k, z) = b^i$, as was adopted in all the fiducial analyses of the first year of DES data. Scale cuts were devised in real space to validate this approximation as well as the Limber approximation \citep{Krause2017}. 
We study the scale cuts in harmonic space in section \ref{sec:scalecuts}.

When comparing with data, the angular power spectrum $C(\ell)$ must be binned in a given set of multipole ranges $\Delta \ell$. This binning and the effect of the mask will be discussed in Section \ref{sec:method}.

The galaxy angular correlation function $w(\theta)$ can be computed from the angular power spectrum in the flat sky approximation as:
\begin{equation}
\label{eq:xi_in_terms_of_Cell}
w(\theta)^{ij} = \int \frac{d \ell \ell}{2 \pi} J_0(\ell \theta) 
 C^{ij}_{\delta_g \delta_g}(\ell)\,
\end{equation}
where $J_0$ is the $0^{th}$ order Bessel function of the first kind.

\section{Estimators for two-point galaxy clustering correlations}
\label{sec:method}

In this section we present the estimators used for the measurements of the two-point galaxy power spectrum and angular correlation functions.
Our starting point is the fluctuation in the number density of galaxies in the direction $\hat{n}$ with respect to the average number density in the observed sample defined as:
\begin{equation}
    \delta_g(\hat{n}) = \frac{n_g(\hat{n}) - \bar{n}_g}{\bar{n}_g}.
\end{equation}
In full sky the fluctuation in the number density in a given position on the sphere $\delta_g(\hat{n})$, can be expanded in spherical harmonics $Y_{\ell m}(\hat{n})$ as
\begin{equation}
\delta_{g}(\hat{n}) = \sum_{\ell,m} a_{\ell m} Y_{\ell m}(\hat{n})
\end{equation}
The angular power spectrum $C_{\ell}$ is
defined as 
\begin{equation}
\langle a_{\ell'm'} a_{\ell m} \rangle = \delta_{\ell', \ell} \delta_{m',m} C_{\ell}.
\end{equation}

The angular power spectrum can be estimated as
\begin{equation}
    \hat{C}_{\ell} = \frac{1}{2 \ell +1} \sum_m |a_{\ell m}|^2.
\end{equation}
However, when the survey does not cover the full sky the procedure above can still be carried out but it would result in the so-called pseudo-$C_{\ell}$, denoted by $\Tilde{C}_{\ell}$. In partial sky the spherical harmonics are not orthogonal anymore and a mixture of $\ell$ modes contributes to the true power spectrum. The survey area is characterized by a mask that gives weights to the different regions of the survey. 
We use the pseudo-$C_\ell$ method developed in \citet{Hivon:2001jp} and implemented in code {\tt  NaMaster} \footnote{\tt github.com/LSSTDESC/NaMaster} \citep{Alonso2019} to recover the true $C_\ell$'s by means of the so-called coupling matrix $M$:
\begin{equation}
    \Tilde{C}_\ell = \sum_{\ell'}  M_{\ell' \ell} C_{\ell'}
\end{equation}
The coupling matrix is solely determined by the survey mask and can be computed numerically in terms of Wigner $3j$ symbols by {\tt NaMaster}.
One needs to invert the coupling matrix or alternatively to forward-model the pseudo-$C_\ell$ to use in a likelihood analyses. We will use the former approach. 

We will model the angular power spectrum with a given binning of $\ell$'s and denote the binned angular power spectrum by $C_q$, where $q$ defines a range of $\ell$'s, $\{\ell_q^1, \ell_q^2, \dotsc ,\ell_q^{n_q}\}$, and $n_q$ is the number of modes grouped in that particular bin. The binned pseudo-$C_\ell$ in a bin $q$ can be written as:
\begin{equation}
    \Tilde{C}_q = \sum_{\ell \in q} w_q^\ell \Tilde{C}_\ell
\end{equation}
where $ w_q^\ell$ is a weight for each $\ell$-mode normalized as $\sum_{\ell \in q} w_q^\ell=1$. Unless otherwise stated we will adopt equal weights.
The binned angular power spectrum is given by:
\begin{equation}
   \Tilde{C}_q = \sum_{q'} {\cal M}_{q q'}  C_{q'}
\end{equation}
where the binned coupling matrix is written as:
\begin{equation}
     {\cal M}_{q q'} =  \sum_{\ell \in q} \sum_{\ell' \in q'}  w_q^\ell M_{\ell' \ell}.
\end{equation}
Inverting the binned coupling matrix is numerically more stable compared to the unbinned matrix.

The theoretical prediction for the binned power spectrum must be corrected as \citep{Alonso2019}
\begin{equation}
    C^{th}_q = \sum_{l} {\cal F}_{q \ell}  C^{th}_\ell,
\end{equation}
where the filter matrix ${\cal F}$ is given by 
\begin{equation}
    {\cal F}_{q \ell} = \sum_{q q'} {\cal M}_{q q'}^{-1} \sum_{\ell'\in q'} w_{q'}^{\ell'} M_{\ell' \ell}.
\end{equation}

We also estimate the two-point angular correlation function $w(\theta$) between the galaxy distribution in 2 directions separated by an angle $\theta$:
\begin{equation}
    w(\theta) = \langle \delta_g(\hat{n}+\hat{\theta}) \delta_g(\hat{n}) \rangle
\end{equation}
where $\theta$ is the angle between directions $\hat{n}$ and $\hat{n} + \hat{\theta}$.

The effect of the partial sky is taken into account by computing correlations of galaxies in the actual catalogue and also from a random catalogue within the survey area. 

We use the  Landy-Szalay estimator \citep{Landy1993} is estimating the galaxy clustering correlation function inside an angular bin $[\theta_{1}, \theta_{2}]$ as
\begin{equation}
\label{eq:Landy_Szalay}
    \hat w[\theta_{1}, \theta_{2}] = \frac{DD[\theta_{1}, \theta_{2}] - 2DR[\theta_{1}, \theta_{2}] + RR[\theta_{1}, \theta_{2}]}{RR[\theta_{1}, \theta_{2}]}\ ,
\end{equation}
where $DD[\theta_{1}, \theta_{2}]$ is the number of galaxy pairs found within the angular bin, $RR[\theta_{1}, \theta_{2}]$ is the (normalised) number of pairs of random points that uniformly samples the survey footprint and $DR[\theta_{1}, \theta_{2}]$ the (normalised) number of galaxy-random-point pairs within the angular bin. If the number density of random points $n_r$ is much larger than the number density of the galaxies $n_g$ (as is recommended for reduce sampling noise) then both $RR$ and $DR$ must be rescaled by factors of $(n_g/n_r)^2$ and $(n_g/n_r)$ respectively.

\begin{figure}
    \centering
    \includegraphics[width=0.5\textwidth]{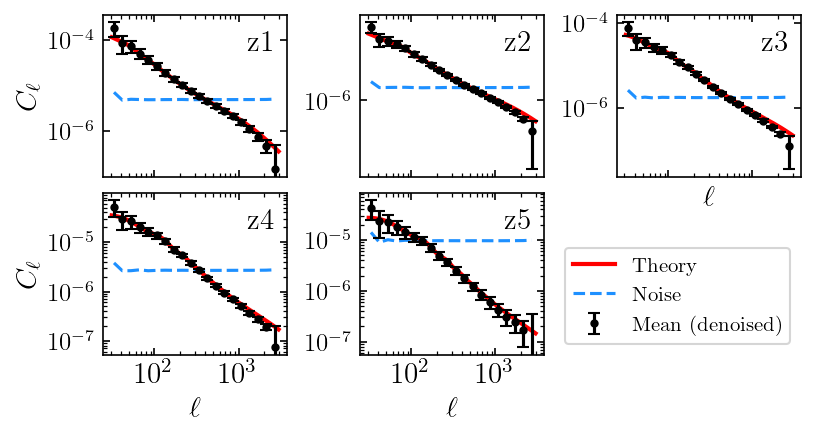}
    \caption{Measured galaxy clustering angular power spectra on the 1200 DES-Y1 FLASK mocks. We show the mean of the noise subtracted measurements along the mock realizations with the 1-$\sigma$ dispersion (black dots), the
    un-binned theoretical input for the realizations (red line) and the estimated binned noise (dot-dashed blue line) for the 5 photometric redshift bins.}
    \label{fig:measurements}
\end{figure}

\section{DES-Y1 mocks}

\subsection{Description of the mocks} 
\label{sec:mocks}

We use a suite of 1200 lognormal simulations generated using the Full-sky Lognormal Astro-fields Simulation Kit (FLASK\footnote{\tt www.astro.iag.usp.br/$\sim$flask}) \citep{Xavier:2016elr} for the DES-Y1 analyses
with resolution Nside=4096.
The input matter angular power spectra were computed using CosmoLike \citep{Krause:2016jvl} with a flat $\Lambda$CDM model with the following parameters 
(which we call FLASK cosmology): 
$\Omega_m$ = 0.285, $\Omega_b$ = 0.05, $\sigma_8=0.82$, $h$ = 0.7, $n_s$ = 0.96 and $\sum m_\nu = 0$.

Galaxy maps were generated in 5 photometric redshift ($z_{ph}$) bins, with ranges 
$(0.15 - 0.3)$,  $(0.3 - 0.45)$, $(0.45 - 0.6)$, $(0.6 - 0.75)$
and  $(0.75 - 0.9)$, and with redshift distributions given in \cite{Elvin-Poole:2017xsf}. The average number density of objects in each tomographic bin are: 0.01337, 0.03434, 0.05094, 0.03297 and 0.00886  arcmin$^{-2}$. We also adopted a linear bias model with a fixed value for each redshift bin given by
$b = 1.45, 1.55, 1.65, 1.8$ and $2.0$ respectively.  

\subsection{ Measurements on DES-Y1 mocks}
\label{sec:mockmeasurement}

\begin{figure}
    \centering
    \includegraphics[width=0.5\textwidth]{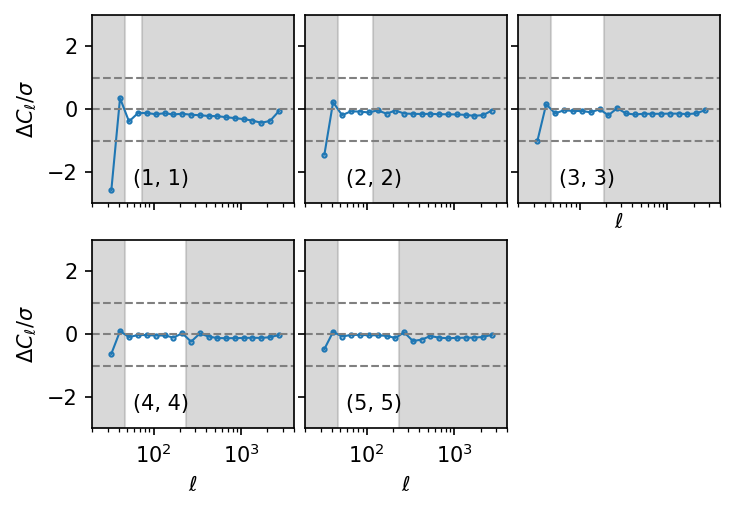}
    \caption{Difference between the interpolated theoretical $C_\ell$ evaluated at the center of the bin and the properly binned theoretical prediction measured in terms of its standard deviation for the 5 redshift bins.The grey regions are the ones excluded by the physical scale cuts (see section \ref{sec:resultsmocks}).}
    \label{fig:binning}
\end{figure}

The measurements of the angular power spectrum for the 1200 mocks were performed using {\tt NaMaster}, as described in section \ref{sec:method}.  We measured pseudo-$C_\ell$'s in $20$ logarithmic bins with $\ell_{min}=30$, $\ell_{max} = 3000$ and a resolution of Nside=2048. The DES-Y1 mask was used to compute the coupling matrix to obtain the binned $C_\ell$'s.

In order to obtain a clean measurement of the clustering signal, we subtracted a noise term assumed to be purely Poissonian. For that, we follow the analytical derivation in~\citet{Alonso2019} for the shot noise in the pseudo-$C_\ell$ measurement, $f_{\rm sky} / \bar{n}_g^i$, where $f_{\rm sky}$ is the covered fraction of the sky, defined by the angular mask, and $\bar{n}_g^i$ the angular density of galaxies in the $i$-th tomographic redshift bin in units of inverse steradians.
Here we use the input average density of galaxies for each redshift bin $i$ for all mocks.

In Figure \ref{fig:measurements} we show the average of the 1200 shot-noise subtracted measurements of the auto (same redshift bin) angular power spectrum and compare them with the un-binned input $C_\ell$'s demonstrating good agreement. 

However, one can notice that the first bin in $\ell$ lies systematically higher than the theoretical input in all 5 redshift bins. 
The reason is that the input $C_\ell$'s are not binned (we show the value of $C_\ell$ at the center of the bin). We show in Figure \ref{fig:binning} that properly taking into account binning in the theoretical input $C_\ell$'s affects only the largest scales and it actually improves the agreement of the first bin.

The measurements of the real-space correlation function $w(\theta)$ used in this work were performed on the same mocks using the code {\tt TreeCorr} \citep{Jarvis2004} with the parameter binslop set to $0.1$ and  the Landy-Szalay estimator \citep{Landy1993}.

\begin{figure}
  \includegraphics[width=0.5\textwidth]{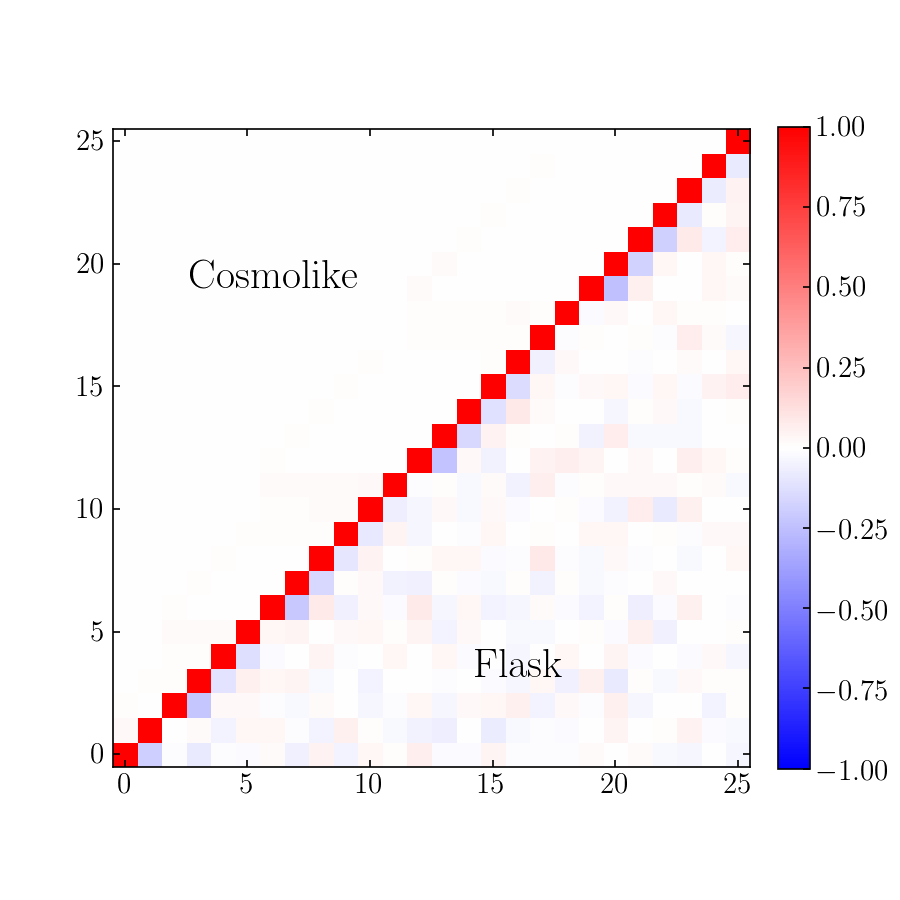}
  \caption{ Correlation matrices from the FLASK mocks (lower right triangle) compared to the analytical matrix from CosmoLike using the FLASK cosmology. }
  \label{fig2}
\end{figure}

\section{Covariance matrix}
\label{sec:covariance}

We use these measurements on the 1200 mocks to estimate a sample covariance matrix for the angular power spectrum, properly taking into account the Hartlap correction factor~\citep{Hartlap2007} for its associated precision matrix.
We compare the FLASK covariance matrix with an analytical one 
generated using CosmoLike with FLASK cosmology which includes nongaussian contributions and the mask is treated using the so-called f$_{\mbox{\small{sky}}}$ approximation.

We also produced a theoretical covariance matrix using an adapted version of the public code CosmoCov\footnote{\url{https://github.com/CosmoLike/CosmoCov}} \citep{2020MNRAS.497.2699F} based on CosmoLike framework \citep{Krause:2016jvl}, with configurations consistent with the Y1 3x2pt analyses. This theoretical covariance includes nongaussian contributions given by the trispectrum and the super-sample covariance.
However, it does not contain the effect of the mode coupling induced by the angular mask and also it does not introduce a bandpower binning.

A comparison of the CosmoLike and Flask correlation matrices is shown in Figure \ref{fig2} (we have also produced a gaussian covariance matrix that takes into account the DES-Y1 survey mask using {\tt NaMaster} obtaining similar results). 
As expected, the FLASK covariance is much noiser than the analytical one.
However, we will show in the next section that the impact of these differences in parameter estimation will be negligible after applying scale cuts.

Given the fact that the FLASK covariance is more realistic, taking into account, by construction, the mode coupling from angular masking and proper bandwidth binning, we will be using it as our default choice.

\section{Analysis pipeline and scale cuts}
\label{sec:resultsmocks}

\subsection{Analysis pipeline}
We have developed a pipeline for estimating parameters from the 
galaxy angular power spectrum based on {\tt CosmoSIS} \citep{2015A&C....12...45Z} using the {\tt MultiNest} sampler \citep{Feroz:2008xx, Feroz:2013hea}. We also use the existing DES-Y1 3x2pt CosmoSIS pipeline for the angular correlation function in order to compare our results. 
In order to explore the parameter space, we compute the Likelihood, defined as:
\begin{equation}
    -2 \ \mathrm{log} \mathcal{L} (\vec{p}) \equiv \chi^2 = \sum_{ij} (D_i - C_{\ell}{}_{i} (\vec{p})) \; \mbox{ Cov}^{-1}_{ij}(D_j - C_{\ell}{}_{j} (\vec{p}) ); 
    \label{chi2}
\end{equation}

\noindent where $C_{\ell}{}_{i}$ is the predicted value at the effective $\ell_{i}$,  $D_i$ is the measured data point and $\mathrm{Cov}$ is the covariance matrix used in the analysis. 
Finally, in the scenarios analysed in this work, the best-fit set of parameters was found using the {\tt MINUIT2} routine \citep{James:1975dr}.

In this Section we test this pipeline for the average of the FLASK realizations and also for one particular realization. We also study the effect of different scale cuts on the angular power spectrum and use our mocks to determine the scale cuts we will use in the DES-Y1 data.

Our goal is to compare our results to previous DES-Y1 results which focused on the estimate of the galaxy bias parameters $b_i$, or more precisely on the amplitude of perturbations given by $b_i \sigma_8$. Therefore, we will also concentrate on these quantities. 

We will run two types of nested sampling chains, depending on the parameters allowed to change: ``quick'' chains with only the 5 bias parameters changing, ``DES-Y1" chains with all 10 nuisance (biases and redshift uncertainties characterized by a shift parameter in the mean of the distribution for each redshift bin). 
All runs in this section adopt the FLASK cosmology described in section  \ref{sec:mocks}.

\subsection{Scale cuts}
\label{sec:scalecuts}
The DES-Y1 analyses defined a scale cut corresponding to a single comoving scale of $R=8\ h^{-1}$Mpc to ensure that the linear bias model does not bias the estimation of the cosmological parameters
in~\citet{Elvin-Poole:2017xsf}. This corresponds to values of $\theta^i_{min} = R/\chi(\langle z^i \rangle)$ given by $43'$, $27'$, $20'$, $16'$ and $14'$ for the five redshift bins with DES-Y1 cosmology. 
The maximum values are set to $\theta_{max} = 250'$ for all bins.

There is no unique and rigorous way to translate the scale cut in real space to harmonic space and we will test two different relations.
We first use a simple relation $\ell = \pi/\theta$ to convert the scale cuts from configuration space to harmonic space and refer to it as ``\textit{naive} scale cuts''. This procedure results in the following values for $\ell_{max}$: 251, 400, 540, 675 and 771 for the 5 redshift bins for DES-Y1 cosmology and $\ell_{min} = 43$ for all bins.

  We also use what we call ``physical scale cuts'', obtained from a hard cut on the comoving Fourier mode
  $k_{\rm max} = 1/R$ related to the minimum comoving scale $R=8\ h^{-1}$ Mpc.
  This cut is translated to an angular harmonic mode on each tomographic bin using the Limber relation $\ell_{\rm max} = k_{\rm max} \times \chi(\langle z^i \rangle)$, where $\langle z^i \rangle = \int z n(z)\, {\rm d}z / \int n(z)\, {\rm d}z$ is the mean redshift for the $i$-th tomographic bin of the analysis and $\chi(z)$ the comoving distance computed on the fiducial cosmology of the analysis.
  Similar approaches were taken in~\cite{2020JCAP...03..044N} and \cite{2020arXiv201106469D}.
  We fix $k_{\rm max} = 0.125\, h\, \mathrm{Mpc}^{-1}$, which yields $\ell_{\rm max}$ = 80, 127, 172, 215, and 246 for the 5 redshift bins for DES-Y1 cosmology.
  We checked these cuts do not change significantly when using a FLASK cosmology.
  We also keep $\ell_{\rm min} = 43$ for all bins based on the impact of binning for our modeling. 
  
 In order to verify whether these scale cuts are effective to mitigate the effects of nonlinear bias we perform a simple $\chi^2$ test comparing two data vectors: a fiducial data vector generated with a linear bias and a data vector contaminated with nonlinear bias. More specifically, we generate a contaminated data vector with an additional quadratic bias parametrized as a function of the linear bias as in \citet{Lazeyras:2015lgp}. Using the FLASK covariance matrix we find that $\Delta \chi^2 =  0.2 $ for the physical cuts, where
 \begin{equation}
     \Delta \chi^2 =  
     \sum_{qq'} (C_q^f - C_q^{b_2}) \; \mbox{ Cov}^{-1}_{q q'} (C_{q'}^f - C_{q'}^{b_2}),
 \end{equation}
 well below the criterion $\Delta \chi^2 <1 $ adopted in \cite{Abbott:2018xao}.
 Hence the nonlinear biases as modelled above are mitigated by the physical scale cuts. Just for comparison, for the naive scale cuts, we find that $\Delta \chi^2 = 14.31$ and therefore this further justifies our fiducial analysis choice for the physical scale cuts.
 
\section{Results on mocks}
\label{sec:pipelineonmocks}

\begin{figure}
\includegraphics[width=0.5\textwidth]{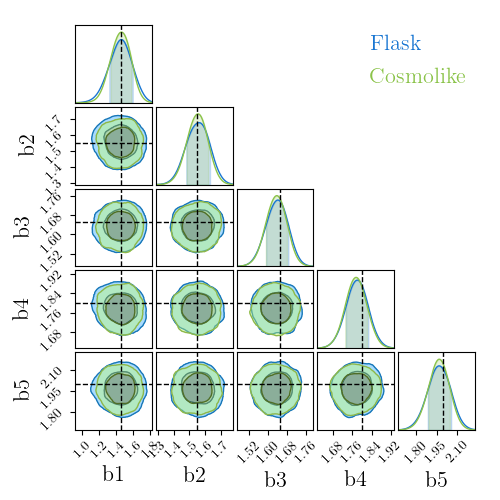}
\caption{Comparison of the bias constraints to the average of the FLASK mocks using either FLASK (DES-Y1 run) or CosmoLike (quick run) covariance matrices with physical scale cuts for the average of the mocks. Inner (outer) contours are drawn at 68\%  (95\%) of confidence level. }
  \label{fig:covcomparison}
\end{figure}

\begin{table*}
  \centering
  \begin{tabular}{ccccccc}
    \hline
    Model & $b_1 \sigma_8$ & $b_2 \sigma_8$ & $b_3 \sigma_8$ & $b_4 \sigma_8$ & $b_5 \sigma_8$ \\ 
    \hline
    FLASK cosmology & 1.189 & 1.271 & 1.353 & 1.476 & 1.640 \\
    \hline
    Config. Space & $1.186^{+0.074}_{-0.075}$ & $1.274^{+0.047}_{-0.049}$ & $1.347^{+0.037}_{-0.039}$ & $1.471^{+0.042}_{-0.043}$ & $1.617^{+0.077}_{-0.080}$ \\ 
    Physical & $1.194^{+0.100}_{-0.120}$ & $1.274^{+0.061}_{-0.059}$ & $1.340\pm 0.037$ & $1.458^{+0.038}_{-0.041}$ & $1.615^{+0.068}_{-0.071}$ \\ 
    Naive & $1.186^{+0.046}_{-0.050}$ & $1.265^{+0.029}_{-0.027}$  & $1.333\pm 0.022$ & $1.454^{+0.027}_{-0.026}$ & $1.602^{+0.058}_{-0.055}$ \\ 
    \hline
  \end{tabular}
  \caption{Measurements of galaxy bias for the different redshift bins from the average of mocks, using Flask covariances in three different cases (i) for the configuration space using the pipeline of this paper,  and for harmonic space (ii) \textit{physical} scale cuts and (iii) the \textit{naive} scale cuts (Figure \ref{fig:average_mocks_2D}). Harmonic space with physical scale cuts and configuration space have a good agreement between each other.  
}
  \label{table:average}
\end{table*}

In order to validate our pipeline, we ran the implemented nested sampling algorithm using as input a random mock realization as well as the average of the set of mocks. We explored the parameter space:
\begin{equation}
\vec{p} = \{b_1,\; b_2,\; b_3,\; b_4,\; b_5,\; \Delta z_1,\; \Delta z_2,\; \Delta z_3,\; \Delta z_4,\; \Delta z_5\; \}
\end{equation}

\noindent where $b_i$ is the constant linear galaxy bias in the redshift shell $z_i$ with a flat prior ($0.8 < b_i <3.0$), and $\Delta z_i$ is the respective shift in the mean of the photometric redshift distribution for each redshift bin $i$: $n_i(z) \rightarrow n_i(z - \Delta z_i)$. The gaussian priors are centered in $\Delta z_i=0$ (for the synthetic data used in this section) and width $\sigma_{\Delta z_i} =\{0.007, 0.007, 0.006,0.010,0.010 \}$ as  in \cite{Elvin-Poole:2017xsf}.

We use the FLASK covariance matrix as the fiducial one. We show that the CosmoLike theoretical covariance matrix yields very similar results in Figure \ref{fig:covcomparison}. 

\begin{figure}
\includegraphics[width=0.5\textwidth]{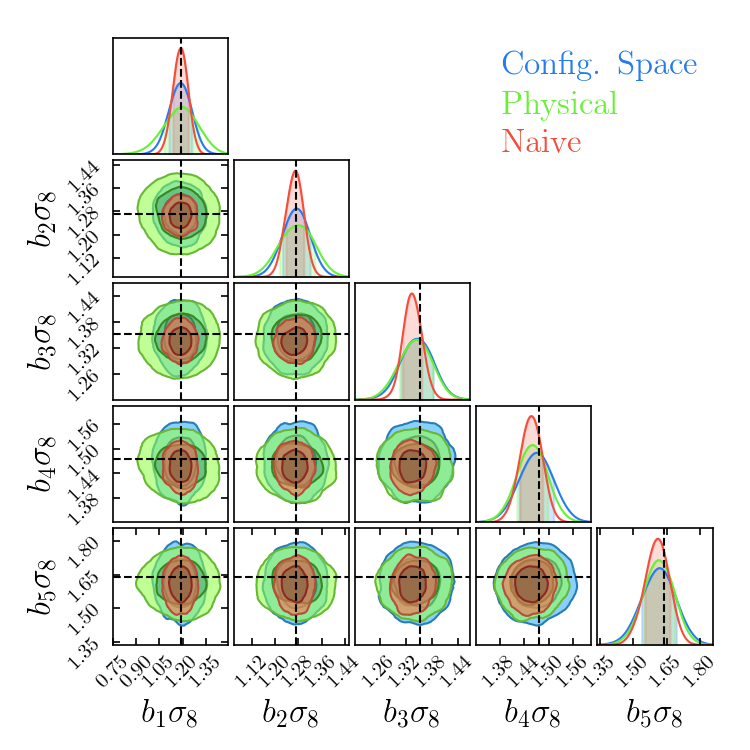}

\caption{Constraints obtained from the analyses on configuration  space  and  harmonic  space  for  the  combination  of  the  5  linear bias $b_i$ and $\sigma_8$, for the average of the FLASK mocks. The green (red) region are constraints using the physical (naive) scale cut (section \ref{sec:scalecuts}).
Inner (outer) contours are drawn at 68\%  (95\%) of confidence level. }
  \label{fig:average_mocks_2D}
\end{figure}

\begin{figure}
\includegraphics[width=0.5\textwidth]{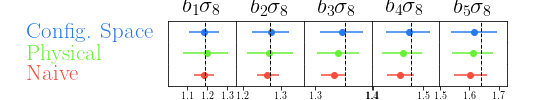}
\caption{Configuration space and harmonic space constraints (for 2 scale cuts) on $b_i  \; \sigma_8$ for the average of the FLASK mocks. The bar represents 68\% of confidence level. }
  \label{fig:summary_mocks_2D}
\end{figure}

Figures \ref{fig:average_mocks_2D} and \ref{fig:summary_mocks_2D} show the parameter countours for the average of the measurements of the angular correlation function $w(\theta)$ using DES-Y1 cuts and the angular power spectrum $C_\ell$ on the mocks. In addition to the good agreement between the estimates from configuration and harmonic spaces 
one can see that the physical scale cuts result in contours more similar to the configuration space (notice that both cuts are consistent with configuration space results at 1 $\sigma$).

\begin{figure}
\includegraphics[width=0.5\textwidth]{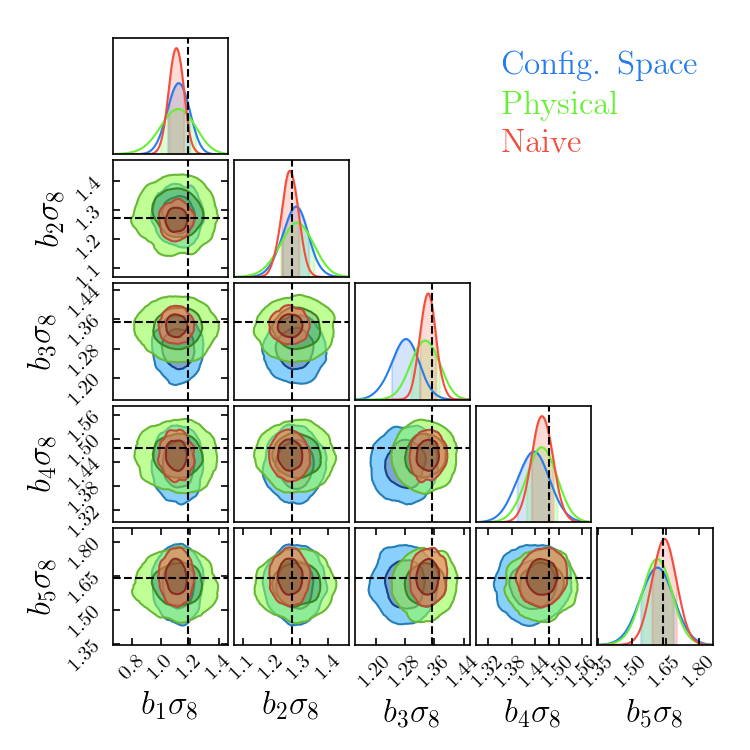}
\caption{Configuration space and harmonic space contours (for 2 scale cuts) on $b_i \sigma_8$ for a single FLASK mock. Inner (outer) contours are drawn at 68\%  (95\%) of confidence level.}
  \label{fig:single_mocks_2D}
\end{figure}

\begin{figure}
\includegraphics[width=0.5\textwidth]{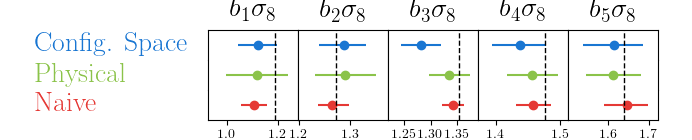}
\caption{Configuration space and harmonic space constraints (for 2 scale cuts) for a single FLASK mock. The bar represents 68\% of confidence level. }
 \label{fig:summary_single_mocks_2D}
\end{figure}

The same general behaviour is seen with the analysis of a single mock, where larger statistical fluctuations are expected, as shown in Figures \ref{fig:single_mocks_2D} and \ref{fig:summary_single_mocks_2D}.
The measurements of the combination $b_i \sigma_8$ considering different scale cuts are given in Table~\ref{table:average}, for the average of mocks as the data vector, and Table~\ref{table:single}, for a single mock realization. In Figures \ref{fig:single_mocks_2D} and \ref{fig:summary_single_mocks_2D} we show that our pipeline is able to recover the input values from a simulated data vector.
Some deviation from the input values are however expected due to statistical fluctuations.

\begin{table*}
  \centering
  \begin{tabular}{cccccc}
    \hline
    Model & $b_1 \sigma_8$ & $b_2 \sigma_8$ & $b_3 \sigma_8$ & $b_4 \sigma_8$ & $b_5 \sigma_8$ \\ 
    \hline
    FLASK cosmology & 1.189 & 1.271 & 1.353 & 1.476 & 1.640 \\
    \hline
    Config. Space & $1.121\pm 0.077$ & $1.287^{+0.044}_{-0.048}$ & $1.282^{+0.038}_{-0.039}$ & $1.437^{+0.040}_{-0.044}$ & $1.613^{+0.073}_{-0.076}$ \\ 
    Physical &$1.133^{+0.120}_{-0.130}$ & $1.287^{+0.061}_{-0.060}$ & $1.337\pm 0.039$ & $1.459^{+0.040}_{-0.039}$ & $1.619^{+0.075}_{-0.069}$ \\ 
    Naive & $1.104^{+0.054}_{-0.049}$ & $1.262^{+0.029}_{-0.028}$ & $1.339^{+0.023}_{-0.021}$ & $1.451^{+0.028}_{-0.026}$ & $1.636\pm 0.055$ \\ 
    \hline
  \end{tabular}
  \caption{
Measurements of galaxy bias for the different redshift bins from a single mock realization, using Flask covariances in three different cases (i) for the configuration space using the pipeline of this paper,  and for harmonic space (ii) \textit{physical} scale cuts and (iii) the \textit{naive} scale cuts (Figure \ref{fig:single_mocks_2D}). Harmonic space with physical scale cuts and configuration space have a good agreement between each other.
  }
  \label{table:single}
\end{table*}

\begin{figure}
\includegraphics[width=0.48\textwidth]{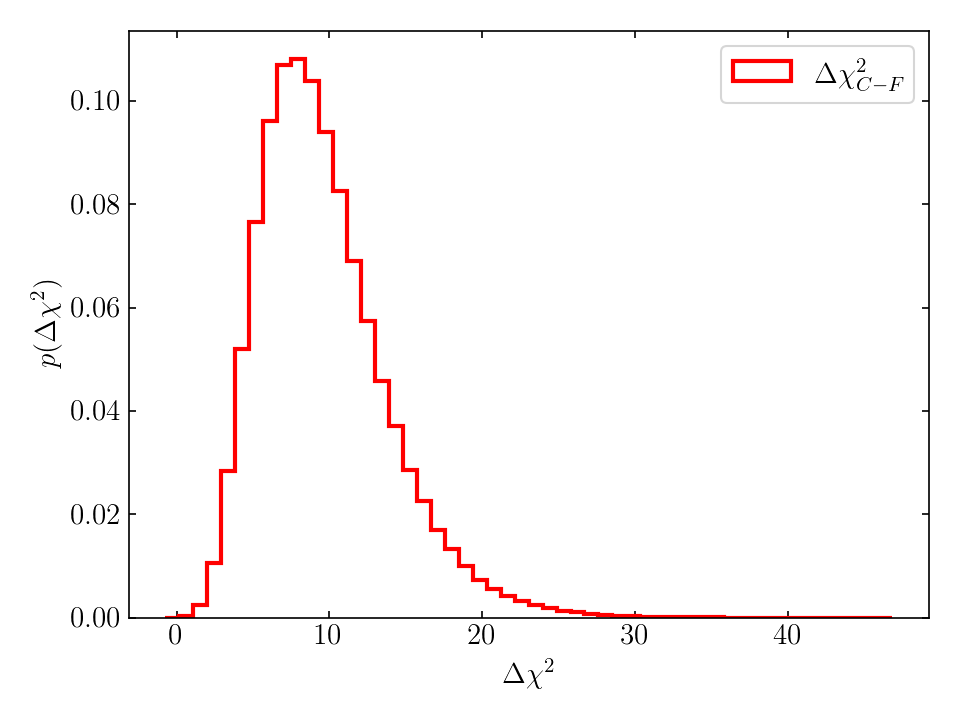}
\caption{Histogram of $\Delta \chi^2$ for 
the CosmoLike covariance and the Flask covariance (fiducial) and lognormal
datavectors. We computed the  $\chi^2$ of mocks by the method described in
section 7 of~\protect\cite{Friedrich:2020dqo}. From the resulting distribution of $\Delta \chi^2=\chi^2_{CosmoLike} - \chi^2_{Flask}$,
we obtain $\Delta \chi^2=  9.63 \pm 4.19$.  This value are in agreement to the theoretical estimate (section \ref{sec:pipelineonmocks}), namely, $\Delta\chi^2_{C-F} = 10.67 \pm 4.30$.}
  \label{fig:shiftchi2}
\end{figure}

As an illustration, for one mock realization, we obtained $\chi^2 = 16.01$
  for 16 degrees of freedom at the best-fit parameters, using the Flask
  covariance matrix and the physical scale cuts. For comparison, the CosmoLike
  covariance matrix yields $\chi^2=20.95$, for the same mock realization and
  scale cuts, giving
  $\Delta\chi^2_{C-F} = \chi^2_{\rm CosmoLike} - \chi^2_{\rm Flask}=4.94$. 
  
This shift can be compared with an analytical approximation for  the expectation value and variance of $\Delta \chi^2_{C-F}$, computed following the prescription in \cite{2020MNRAS.497.2699F},
  \begin{align*}
    E[\Delta\chi^2_{C-F}] &= {\rm Tr}(\mathbf{C}_C^{-1}\mathbf{C}_F) - N_D~,\\
    {\rm Var}[\Delta\chi^2_{C-F}] &= 2N_D + 2 {\rm Tr}(\mathbf{C}_C^{-1}\mathbf{C}_F\mathbf{C}_C^{-1}\mathbf{C}_F)-4{\rm Tr}(\mathbf{C}_C^{-1}\mathbf{C}_F)~.
  \end{align*}
  Here $\mathbf{C}_{C(F)}$ is the CosmoLike (Flask) covariance and $N_D$ the
  number of degrees of freedom,
  which yields $\Delta\chi^2_{C-F} = 10.67 \pm 4.30$. Hence, the
  $\Delta \chi^2$ for the particular realization we choose, has a typical
  deviation from the expected value. Finally, considering the approach in section 7 of  \cite{Friedrich:2020dqo}, we computed the $\Delta \chi^2$ distribution from our set of 1200 mocks, obtaining  $\Delta \chi^2=  9.63 \pm 4.19$ (see Figure \ref{fig:shiftchi2}). These values corroborate the the analytical estimates of $\chi^2$ shift from the different covariance matrices. Detailed investigations of the survey mask effects are discussed in  \cite{Friedrich:2020dqo}.

We notice that the more aggressive physical scale cuts in harmonic space yielded less biased results for the single mock analysis and is more compatible with the real-space analysis. 
Therefore we opt to choose the physical scale cuts when studying the compatibility between the configuration and harmonic space analyses in the DES-Y1 data in the next Section.

\section{DES-Y1 results on bias from galaxy clustering in harmonic space}
\label{sec:resultsdata}
We use data taken in the first year (Y1) of DES observations (DES-Y1) \citep{Diehl:2014lea}.
In particular, we will follow very closely \citet{Elvin-Poole:2017xsf} and use the catalogue, redshift distributions and systematic weights obtained in the configuration space galaxy clustering analysis performed for the DES-Y1 data.

\subsection{Data}
\label{sec:data}
We use the catalogue created in \citet{Elvin-Poole:2017xsf}, where details can be found, built from the original so-called ``DES-Y1 Gold'' catalogue \citep{Drlica-Wagner:2017tkk}, with an area of approximately 1500 deg$^2$.
The galaxy sample in \citet{Elvin-Poole:2017xsf} was generated
by the redMaGiC algorithm \citep{Rozo_2016}, run on DES-Y1 Gold
data. 
The redMaGiC algorithm selects Luminous Red Galaxies (LRGs) in such a way that photometric redshift uncertainties are minimized, and it produces a luminosity-thresholded sample of constant co-moving density.
The redMaGiC samples were split into five redshift bins
of width $\Delta z = 0.15$ from $z = 0.15$ to $z = 0.9$.
After masking and additional cuts, the total number of objects is
653,691 distributed over an area of 1321 deg$^2$.

\begin{table*}
    \centering
    \begin{tabular}{cccccc}
        \hline
		Model & $b_1 \sigma_8$ & $b_2 \sigma_8$ & $b_3 \sigma_8$ & $b_4 \sigma_8$ & $b_5 \sigma_8$ \\ 
		\hline
		Config. Space & $1.160^{+0.070}_{-0.069}$ & $1.325^{+0.042}_{-0.047}$ & $1.327\pm 0.035$ & $1.593^{+0.041}_{-0.042}$ & $1.639^{+0.057}_{-0.063}$ \\ 
		Harmonic Space & $1.147^{+0.100}_{-0.130}$ & $1.319^{+0.053}_{-0.054}$ & $1.314^{+0.036}_{-0.038}$ & $1.618^{+0.032}_{-0.030}$ & $1.677^{+0.049}_{-0.048}$ \\ 
		\hline
    \end{tabular}
    \caption{Measurements of galaxy bias for the different redshift bins from DES-Y1 data (i) for the configuration space using the pipeline of this paper,  and (ii) for harmonic space with \textit{physical} scale cuts (Figure \ref{fig:des10params2d}). Harmonic space with physical scale cuts and configuration space have compatible measurements.
    }
    \label{table:DESY1}
\end{table*}

\subsection{Survey property maps}
The number density of galaxies observed is affected by the conditions of the survey, which are described by the survey property maps. DES has produced 21 of these maps for the Y1 season, such as depth, seeing, airmass, etc.
The correlation between the galaxy density maps with the survey property maps are a sign of contamination. A so-called weight method was used in \cite{Elvin-Poole:2017xsf} in which weights were applied to the galaxy maps in order to decorrelate them from the survey property maps at the $2-\sigma$ level.
We use the weights obtained in \cite{Elvin-Poole:2017xsf} in our analysis by correcting the pixelized number counts map of galaxies with these weights as multiplicative corrections.

\subsection{Angular power spectrum measurements and results in DES-Y1}

We follow the analysis described in Sections \ref{sec:mockmeasurement} and \ref{sec:resultsmocks} validated on the lognormal FLASK mocks. We use the FLASK covariance for the angular power spectrum results. We show results from nested sampling DES-Y1 runs using the {\tt MultiNest} sampler \citep{Feroz:2008xx} for ten parameters (5 galaxy biases and 5 redshift shifts), keeping the other parameters fixed at DES-Y1 cosmology.

\begin{figure}
\includegraphics[width=0.5\textwidth]{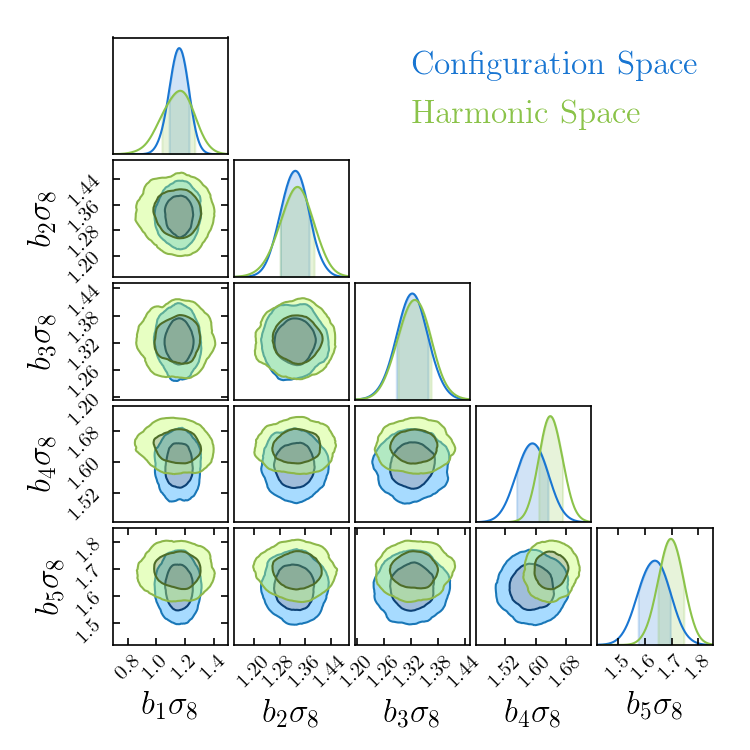}
\caption{Comparison on the constraints obtained from the analyses on  configuration space and harmonic space for the combination of the 5 linear bias $b_i$ and $\sigma_8$, for DES-Y1 data. We are using physical cuts for the angular power spectrum. Inner (outer) contours are drawn at 68\%  (95\%) of confidence level. }
  \label{fig:des10params2d}
\end{figure}

\begin{figure}
\includegraphics[width=0.5\textwidth]{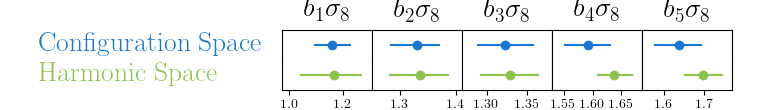}
\caption{Marginalized results in configuration space and harmonic on $b \sigma_8$  for DES-Y1 data. The error bars represents 68\% of confidence level. }
  \label{fig:des10params1d}
\end{figure}

Finally, we present our main results in Figures \ref{fig:des10params2d} and \ref{fig:des10params1d}  from a DES-Y1 run applying the physical scale cuts to the angular power spectrum and compare them with results from a similar run in configuration space.  

\begin{figure}
\includegraphics[width=0.5\textwidth]{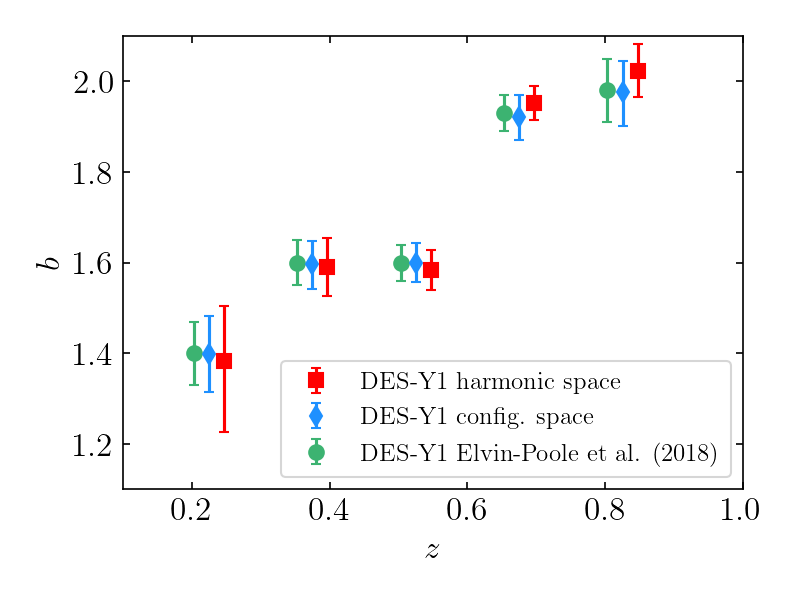}
\caption{Comparison of the results on galaxy bias for the different redshift bins from our analyses in configuration space and harmonic space and the results from \citet{Elvin-Poole:2017xsf} for DES-Y1 data also in configuration space. The error bars represents 68\% of confidence level. }
  \label{fig:comparison}
\end{figure}

We also present a comparison of the results for galaxy biases for the different redshift bins from our analyses in configuration space and harmonic space with the results from \citet{Elvin-Poole:2017xsf} in Figure \ref{fig:comparison}.

\begin{figure}
\includegraphics[width=0.5\textwidth]{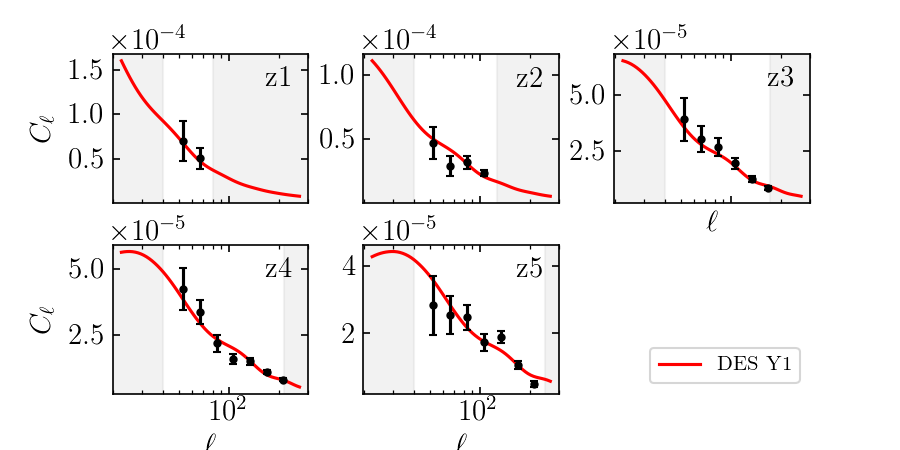}
\caption{Measurements of angular power spectrum in DES-Y1 compared to a model with the  best fit values for galaxy bias. Grey bands are the physical scale cuts. }
  \label{fig:Cl_bestfit}
\end{figure}

We show in Figure \ref{fig:Cl_bestfit} the DES-Y1 angular power spectrum measurements including the scale cuts compared to the best fit predictions.

Fixing the cosmological parameters, we measured the galaxy biases for the harmonic space analysis (with physical scale cut), as shown in Table \ref{table:DESY1}. In order to compare the analyses in harmonic and configuration space, we also used our pipeline to constrain the same set of parameters in configuration space in Figure \ref{fig:des10params2d}.

We analyse the goodness-of-fit by computing the reduced $\chi^2$ (Eq \ref{chi2}) at best-fit parameters.
For the DES-Y1 data in harmonic space, we obtain $\chi^2=28.2$ at the best-fit set of parameters $\vec{p}$  for $\nu=26 -10$ degrees of freedom.
These results yield to a probability to exceed (PTE) of 3.0\%. Nevertheless, as remarked in  \cite{Elvin-Poole:2017xsf}, since the five nuisance parameters $\Delta z_i$ are strongly prior dominated, one can consider here the effective number only the remaining five parameters as the free parameters, which leads in our case to $\nu=26-5$ and a PTE of 13.5\%.
The quoted values of PTE shows that the pipeline was able to produce a reasonable fit to the DES-Y1 data in harmonic space. 
The probability to exceed found by \cite{Elvin-Poole:2017xsf} are 1.4\% and 4.5\% for 10 and 5 free parameters, respectively, and 54 data points. These values are compatible  to what we have found in this work. 

\section{Conclusions}
\label{sec:conclusions}

In this work we present a pipeline to estimate cosmological parameters
from tomographic measurements of the angular power spectrum of galaxy clustering in the DES-Y1 data within the framework of {\tt CosmoSIS} and using several products originally developed by the DES collaboration for the real-space analysis. 
We focus on the determination of the linear galaxy bias in order to compare with the DES-Y1 analysis of the full-shape angular correlation function in real-space.

We tested the pipeline in a suite of lognormal simulations, devise scale cuts in harmonic space and applied the analysis to DES-Y1 data. 
We showed that different covariance matrices from FLASK and CosmoLike produce similar constraints.
Our analysis makes use of the sharp scale cuts devised for the DES-Y1 real-space analysis adapted to harmonic space. Sharp cuts in configuration space do not map exactly into sharp cuts in harmonic space and we study two possibilities which we call naive and physical scale cuts. We conclude that the latter produce results that are more consistent with the real-space analysis on the mocks. Finally, applying our pipeline to DES-Y1 data We find that our results are consistent with the DES-Y1 real-space analysis of \citet{Elvin-Poole:2017xsf}.

We are currently working on a complete 3x2pt analyses in harmonic space for the DES-Y1 data and plan to do the same for DES-Y3 data \footnote{After the completion of this work, a preprint appeared with an analysis of DES-Y1 public data in harmonic space with a model for galaxy bias based on Effective Field Theory \citep{Hadzhiyska:2021xbv} more sophisticated than the fiducial one used in DES-Y1.}.
These results are an initial step towards a multi-probe analyses using the angular power spectra for the Dark Energy Survey with a goal of demonstrating their compatibility and eventually combining real and harmonic space results.

\section*{Acknowledgements}

We sincerely thank Henrique Xavier for helpful discussions on lognormal mock simulations and the FLASK code. 

This research was partially supported by the Laborat\'orio Interinstitucional de e-Astronomia (LIneA), the Brazilian funding agencies  CNPq and CAPES,
the  Instituto Nacional de Ci\^{e}ncia e Tecnologia (INCT) e-Universe (CNPq grant 465376/2014-2) and the Sao Paulo State Research Agency (FAPESP) through grants 2019/04881-8 (HC) and 2017/05549-1 (AT).
The authors acknowledge the use of computational resources from LIneA, the Center for Scientific Computing (NCC/GridUNESP) of the Sao Paulo State University (UNESP), and from the National Laboratory for Scientific Computing (LNCC/MCTI, Brazil), where the SDumont supercomputer ({\tt sdumont.lncc.br}) was used.
This research used resources of the National Energy Research Scientific Computing Center (NERSC), a U.S. Department of Energy Office of Science User Facility operated under Contract No. DE-AC02-05CH11231.

This paper has gone through internal review by the DES collaboration. Funding for the DES Projects has been provided by the U.S. Department of Energy, the U.S. National Science Foundation, the Ministry of Science and Education of Spain, 
the Science and Technology Facilities Council of the United Kingdom, the Higher Education Funding Council for England, the National Center for Supercomputing 
Applications at the University of Illinois at Urbana-Champaign, the Kavli Institute of Cosmological Physics at the University of Chicago, 
the Center for Cosmology and Astro-Particle Physics at the Ohio State University,
the Mitchell Institute for Fundamental Physics and Astronomy at Texas A\&M University, Financiadora de Estudos e Projetos, 
Funda{\c c}{\~a}o Carlos Chagas Filho de Amparo {\`a} Pesquisa do Estado do Rio de Janeiro, Conselho Nacional de Desenvolvimento Cient{\'i}fico e Tecnol{\'o}gico and 
the Minist{\'e}rio da Ci{\^e}ncia, Tecnologia e Inova{\c c}{\~a}o, the Deutsche Forschungsgemeinschaft and the Collaborating Institutions in the Dark Energy Survey. 

The Collaborating Institutions are Argonne National Laboratory, the University of California at Santa Cruz, the University of Cambridge, Centro de Investigaciones Energ{\'e}ticas, 
Medioambientales y Tecnol{\'o}gicas-Madrid, the University of Chicago, University College London, the DES-Brazil Consortium, the University of Edinburgh, 
the Eidgen{\"o}ssische Technische Hochschule (ETH) Z{\"u}rich, 
Fermi National Accelerator Laboratory, the University of Illinois at Urbana-Champaign, the Institut de Ci{\`e}ncies de l'Espai (IEEC/CSIC), 
the Institut de F{\'i}sica d'Altes Energies, Lawrence Berkeley National Laboratory, the Ludwig-Maximilians Universit{\"a}t M{\"u}nchen and the associated Excellence Cluster Universe, 
the University of Michigan, the National Optical Astronomy Observatory, the University of Nottingham, The Ohio State University, the University of Pennsylvania, the University of Portsmouth, 
SLAC National Accelerator Laboratory, Stanford University, the University of Sussex, Texas A\&M University, and the OzDES Membership Consortium.

Based in part on observations at Cerro Tololo Inter-American Observatory at NSF's NOIRLab (NOIRLab Prop. ID 2012B-0001; PI: J. Frieman), which is managed by the Association of Universities for Research in Astronomy (AURA) under a cooperative agreement with the National Science Foundation.

The DES data management system is supported by the National Science Foundation under Grant Numbers AST-1138766 and AST-1536171.
The DES participants from Spanish institutions are partially supported by MINECO under grants AYA2015-71825, ESP2015-66861, FPA2015-68048, SEV-2016-0588, SEV-2016-0597, and MDM-2015-0509, 
some of which include ERDF funds from the European Union. IFAE is partially funded by the CERCA program of the Generalitat de Catalunya.
Research leading to these results has received funding from the European Research
Council under the European Union's Seventh Framework Program (FP7/2007-2013) including ERC grant agreements 240672, 291329, and 306478.

This manuscript has been authored by Fermi Research Alliance, LLC under Contract No. DE-AC02-07CH11359 with the U.S. Department of Energy, Office of Science, Office of High Energy Physics.

This work made use of the software packages  {\tt ChainConsumer} \citep{chainconsumer}, {\tt matplotlib} \citep{matplotlib}, and {\tt numpy} \citep{numpy}.


\bibliographystyle{mnras}
\bibliography{literature}

\section*{Author Affiliations}
\label{app:affiliations}
$^{1}$ Instituto de F\'{i}sica Te\'orica, Universidade Estadual Paulista, S\~ao Paulo, Brazil\\
$^{2}$ Laborat\'orio Interinstitucional de e-Astronomia - LIneA, Rua Gal. Jos\'e Cristino 77, Rio de Janeiro, RJ - 20921-400, Brazil\\
$^{3}$ Departamento de F\'isica Matem\'atica, Instituto de F\'isica, Universidade de S\~ao Paulo, CP 66318, S\~ao Paulo, SP, 05314-970, Brazil\\
$^{4}$ ICTP South American Institute for Fundamental Research\\ Instituto de F\'{\i}sica Te\'orica, Universidade Estadual Paulista, S\~ao Paulo, Brazil\\
$^{5}$ Department of Physics, University of Michigan, Ann Arbor, MI 48109, USA\\
$^{6}$ Department of Physics and Astronomy, University of Pennsylvania, Philadelphia, PA 19104, USA\\
$^{7}$ Center for Cosmology and Astro-Particle Physics, The Ohio State University, Columbus, OH 43210, USA\\
$^{8}$ Department of Physics, The Ohio State University, Columbus, OH 43210, USA\\
$^{9}$ Department of Astronomy/Steward Observatory, University of Arizona, 933 North Cherry Avenue, Tucson, AZ 85721-0065, USA\\
$^{10}$ Department of Physics, Stanford University, 382 Via Pueblo Mall, Stanford, CA 94305, USA\\
$^{11}$ Kavli Institute for Particle Astrophysics \& Cosmology, P. O. Box 2450, Stanford University, Stanford, CA 94305, USA\\
$^{12}$ Institut d'Estudis Espacials de Catalunya (IEEC), 08034 Barcelona, Spain\\
$^{13}$ Institute of Space Sciences (ICE, CSIC),  Campus UAB, Carrer de Can Magrans, s/n,  08193 Barcelona, Spain\\
$^{14}$ Fermi National Accelerator Laboratory, P. O. Box 500, Batavia, IL 60510, USA\\
$^{15}$ Instituto de Fisica Teorica UAM/CSIC, Universidad Autonoma de Madrid, 28049 Madrid, Spain\\
$^{16}$ CNRS, UMR 7095, Institut d'Astrophysique de Paris, F-75014, Paris, France\\
$^{17}$ Sorbonne Universit\'es, UPMC Univ Paris 06, UMR 7095, Institut d'Astrophysique de Paris, F-75014, Paris, France\\
$^{18}$ Department of Physics \& Astronomy, University College London, Gower Street, London, WC1E 6BT, UK\\
$^{19}$ SLAC National Accelerator Laboratory, Menlo Park, CA 94025, USA\\
$^{20}$ Center for Astrophysical Surveys, National Center for Supercomputing Applications, 1205 West Clark St., Urbana, IL 61801, USA\\
$^{21}$ Department of Astronomy, University of Illinois at Urbana-Champaign, 1002 W. Green Street, Urbana, IL 61801, USA\\
$^{22}$ Institut de F\'{\i}sica d'Altes Energies (IFAE), The Barcelona Institute of Science and Technology, Campus UAB, 08193 Bellaterra (Barcelona) Spain\\
$^{23}$ Physics Department, 2320 Chamberlin Hall, University of Wisconsin-Madison, 1150 University Avenue Madison, WI  53706-1390\\
$^{24}$ Department of Astronomy and Astrophysics, University of Chicago, Chicago, IL 60637, USA\\
$^{25}$ Kavli Institute for Cosmological Physics, University of Chicago, Chicago, IL 60637, USA\\
$^{26}$ Astronomy Unit, Department of Physics, University of Trieste, via Tiepolo 11, I-34131 Trieste, Italy\\
$^{27}$ INAF-Osservatorio Astronomico di Trieste, via G. B. Tiepolo 11, I-34143 Trieste, Italy\\
$^{28}$ Institute for Fundamental Physics of the Universe, Via Beirut 2, 34014 Trieste, Italy\\
$^{29}$ Observat\'orio Nacional, Rua Gal. Jos\'e Cristino 77, Rio de Janeiro, RJ - 20921-400, Brazil\\
$^{30}$ Department of Physics, IIT Hyderabad, Kandi, Telangana 502285, India\\
$^{31}$ Santa Cruz Institute for Particle Physics, Santa Cruz, CA 95064, USA\\
$^{32}$ Department of Astronomy, University of Michigan, Ann Arbor, MI 48109, USA\\
$^{33}$ Institute of Theoretical Astrophysics, University of Oslo. P.O. Box 1029 Blindern, NO-0315 Oslo, Norway\\
$^{34}$ School of Mathematics and Physics, University of Queensland,  Brisbane, QLD 4072, Australia\\
$^{35}$ Center for Astrophysics $\vert$ Harvard \& Smithsonian, 60 Garden Street, Cambridge, MA 02138, USA\\
$^{36}$ Department of Applied Mathematics and Theoretical Physics, University of Cambridge, Cambridge CB3 0WA, UK\\
$^{37}$ Department of Astrophysical Sciences, Princeton University, Peyton Hall, Princeton, NJ 08544, USA\\
$^{38}$ Instituci\'o Catalana de Recerca i Estudis Avan\c{c}ats, E-08010 Barcelona, Spain\\
$^{39}$ Institute of Astronomy, University of Cambridge, Madingley Road, Cambridge CB3 0HA, UK\\
$^{40}$ Centro de Investigaciones Energ\'eticas, Medioambientales y Tecnol\'ogicas (CIEMAT), Madrid, Spain\\
$^{41}$ School of Physics and Astronomy, University of Southampton,  Southampton, SO17 1BJ, UK\\
$^{42}$ Computer Science and Mathematics Division, Oak Ridge National Laboratory, Oak Ridge, TN 37831\\

\end{document}